\documentclass[aps,prapplied,reprint,superscriptaddress]{revtex4-1} 

\usepackage{amsmath,amssymb}
\usepackage{graphicx}
\usepackage{dcolumn}
\usepackage{bm}

\usepackage[utf8]{inputenc}
\usepackage[T1]{fontenc}
\usepackage{mathptmx}
\usepackage{epstopdf,xcolor}
\usepackage{appendix}

\begin{document}


 \title{Linear {Dependence of} Post-irradiation Input Bias Currents {on Pre-irradiation Values} in Silicon Bipolar Microcircuits}

\author{Yu Song}
\email[Corresponding author: ]{kwungyusung@gmail.com} 
\affiliation{College of Physics and Electronic Information Engineering, Neijiang Normal University, Neijiang 641112, China}
\affiliation{Microsystem and Terahertz Research Center \& Institute of Electronic Engineering, China Academy of Engineering Physics, Chengdu 610200, China}

\author{Jie Zhao}
\affiliation{Center for Circuits and Systems, Peng Cheng Laboratory, Shenzhen 518055, China}
\affiliation{Microsystem and Terahertz Research Center \& Institute of Electronic Engineering, China Academy of Engineering Physics, Chengdu 610200, China}

\author{Shun Li}
\affiliation{Microsystem and Terahertz Research Center \& Institute of Electronic Engineering, China Academy of Engineering Physics, Chengdu 610200, China}

\author{Shi-Yao Hou}
\affiliation{Microsystem and Terahertz Research Center \& Institute of Electronic Engineering, China Academy of Engineering Physics, Chengdu 610200, China}
\affiliation{College of Physics and Electronic Engineering \& Center for Computational Sciences, Sichuan Normal University, Chengdu 610068, China}

\author{Hang Zhou}
\affiliation{Microsystem and Terahertz Research Center \& Institute of Electronic Engineering, China Academy of Engineering Physics, Chengdu 610200, China}

\author{Yang Liu}
\affiliation{Microsystem and Terahertz Research Center \& Institute of Electronic Engineering, China Academy of Engineering Physics, Chengdu 610200, China}

\author{Ying Zhang}
\affiliation{Microsystem and Terahertz Research Center \& Institute of Electronic Engineering, China Academy of Engineering Physics, Chengdu 610200, China}

\author{Dechao Meng}
\affiliation{Microsystem and Terahertz Research Center \& Institute of Electronic Engineering, China Academy of Engineering Physics, Chengdu 610200, China}

\author{Gang Dai}
\affiliation{Microsystem and Terahertz Research Center \& Institute of Electronic Engineering, China Academy of Engineering Physics, Chengdu 610200, China}

\author{Jian Zhang}
\email[Corresponding author: ]{jianzhang@uestc.edu.cn}
\affiliation{School of Electronic Science and Engineering, University of Electronic Science and Technology of China, Chengdu 611731, China}

\date{\today}

\begin{abstract}
We find in experiments a linear {dependence of} ionization irradiation-induced degradations on pre-irradiation values of the input bias current in bipolar devices with simple input stages.The {dependence} is found to generally exist in all studied cases of different device types and different irradiation conditions.
A unique behavior of the energy distribution of the interface states ($D_{it}$) under irradiation is suggested as the origin of the observed phenomenon: the generation of interface traps through the depassivation of Si-H bonds located 
near the pre-irradiation interface traps displays a $D_{it}$ as an enlarge of the initial $D_{it}$ and results in the general {linear dependence}.
A more accurate damage prediction method by using the pre-irradiation values is proposed based on the observed phenomenon.
\end{abstract}

\maketitle

\section{Introduction}

Linear bipolar microcircuits, such as operational amplifiers and voltage comparators, are widely used in space applications.
Under persistent ionizing irradiations from the space, the electric properties of these devices will degrade, with the input bias current (IBC) usually as the most sensitive parameter.
The variability of the total ionizing dose (TID) degradation of these devices is an important yet quite unexplored field.
Manufacturer-to-manufacturer, lot-to-lot, wafer-to-wafer, part-to-part, and channel-to-channel response variabilities
have been studied for several bipolar devices~\cite{Blair1963_IEEETNS10-35,
Krieg2001_2001IEEENSREC-167,
Pease1996_IEEEREDW-28,
Johnston1979_IEEETNS26-4769,
Barnaby2000Origins, 
Kruckmeyer2008_2008IEEEREDW-110,
Kruckmeyer2009_2009ECRECS-586,
Ladbury2009Statistical, 
Xapsos2017Inclusion,Elushov2016Taking,
Gorelick2004_IEEETNS51-3679,
Guillermin2016_RADECS2016-1,
Bozovich2017_2017IEEEREDW-1,
song2020defect,song2020origin}.
Large manufacturer variability was found for voltage comparators LM111, LM211, and LM311.
The total dose degradation of the IBC varies by a factor of 100 among manufacturers,
as the physical layout of these transistors is different for each vendor~\cite{Krieg2001_2001IEEENSREC-167}.
The response variability of 108A-type operational amplifier was compared at the diffusion lot, wafer, and sub-wafer levels
for breakout transistors as well as complete circuits. It was found that the lower specification devices from the same wafer
or diffusion lot could be used as test samples to determine the hardness of the low-yield 108A devices~\cite{Johnston1979_IEEETNS26-4769}.
{A strong part-to-part response variation is found in a voltage comparator LM111, for which the low doping of the substrate PNP input transistors was identified as the dominating mechanism~\cite{Barnaby2000Origins}.
Besides, an operational amplifier LM124 was found to exhibit a much weaker part-to-part response variation, 
which was attributed to the difference in the radiation-induced oxide defect build-up and circuit effects in this device~\cite{Barnaby2000Origins}.}
Test of amplifiers LM124, LM111, and LM119, as well as comparators LM139 and LM193 from National Semiconductor reveals that the channel-to-channel variability was minor compared with the unit-to-unit or the wafer-to-wafer variation~\cite{Kruckmeyer2008_2008IEEEREDW-110,Kruckmeyer2009_2009ECRECS-586}.
{The influence of dataset size and statistical model on the bounding estimates of TID degradation was examined and a method for selecting the model with greatest predictive power was developed~\cite{Ladbury2009Statistical}. Recently, inclusion of radiation environment variability in TID hardness assurance methodology has also been discussed~\cite{Xapsos2017Inclusion,Elushov2016Taking}.}
Besides, it is found that the lot-to-lot variability of amplifiers LM111, LM124, OP-27, OP-484, RH1014, and RH1056 can be altered by prior neutron irradiations on the samples~\cite{Gorelick2004_IEEETNS51-3679}.

Even for bipolar devices from the same manufacturer and a single lot, the variability of the degradation can stem from the variations of many factors due to the 
uncertainties in photo/etching, ion implantation, oxidation, passivation, packaging, 
and burn-in etc.
These factors include the layer thickness~\cite{Fleetwood1994_IEEETED41-1953}, layer quality (concentration of oxygen vacancies)~\cite{boch2006dose,boch2006temperature},
hydrogen concentration~\cite{pease2008effects,adell2009irradiation,Fleetwood2002_MR42-523}, and electric field~\cite{fleetwood1994physical,fleetwood1996Radiation} in the base oxide,
which determine the concentrations of oxide trapped charges ($N_{ot}$) and interface traps ($N_{it}$) generated from the irradiation. 
 The factors also include
the structure and doping parameters of the transistors~\cite{Wu1997_IEEETNS44-1914,Barnaby2000Origins}
as well as the operating point bias of the microcircuit~\cite{Barnaby1999_IEEETNS46-1666,Barnaby2000Origins}, 
which determine the contribution efficiency of the ionization defects to the base current and IBC. 
 We notice that, the variations of these factors can also result in the variability of the pre-irradiation IBC. 
Hence, it is natural to ask if there is a {clear dependence of the post-irradiation IBCs on the  
pre-irradiation values} in the bipolar devices.
However, to our best acknowledge, there is barely such investigations.

In this work, we answer the question by carrying out systematic irradiation experiments
on large-sample-size bipolar devices from the same manufacturer and the same lot.
Three typical commercial off-the-shelf (COTS) bipolar devices with simple input stages
are chosen as the research subjects.
Remarkably, we find a general {linear dependence of the irradiation response on} the initial values of the IBC, which maintains for all investigated cases: three kinds of devices of different types and
many irradiation conditions of different doses, dose rates, and temperatures.
{We attribute this general dependence to a unique response of the energy distribution of interface traps ($D_{it}$) under irradiation: the generation of interface traps near the inital ones results in the enlarge of $D_{it}$, which demonstrates as a linearly-dependent component in the post-irradiation IBC; the generation of interface traps away from the initial ones results in the overall shift of $D_{it}$, which displays as an independent component in the post-irradiation IBC.}
Based on the observed phenomenon, we examine
the applicability of the traditional mean-value damage prediction method
and propose a more accurate method for bipolar devices {whose post-irradiation IBCs display strong dependence on the pre-irradiation values}.

\section{Experimental Setup}
\label{sec:setup}

\begin{figure}[!t]
\centering
\includegraphics[width=0.89\linewidth]{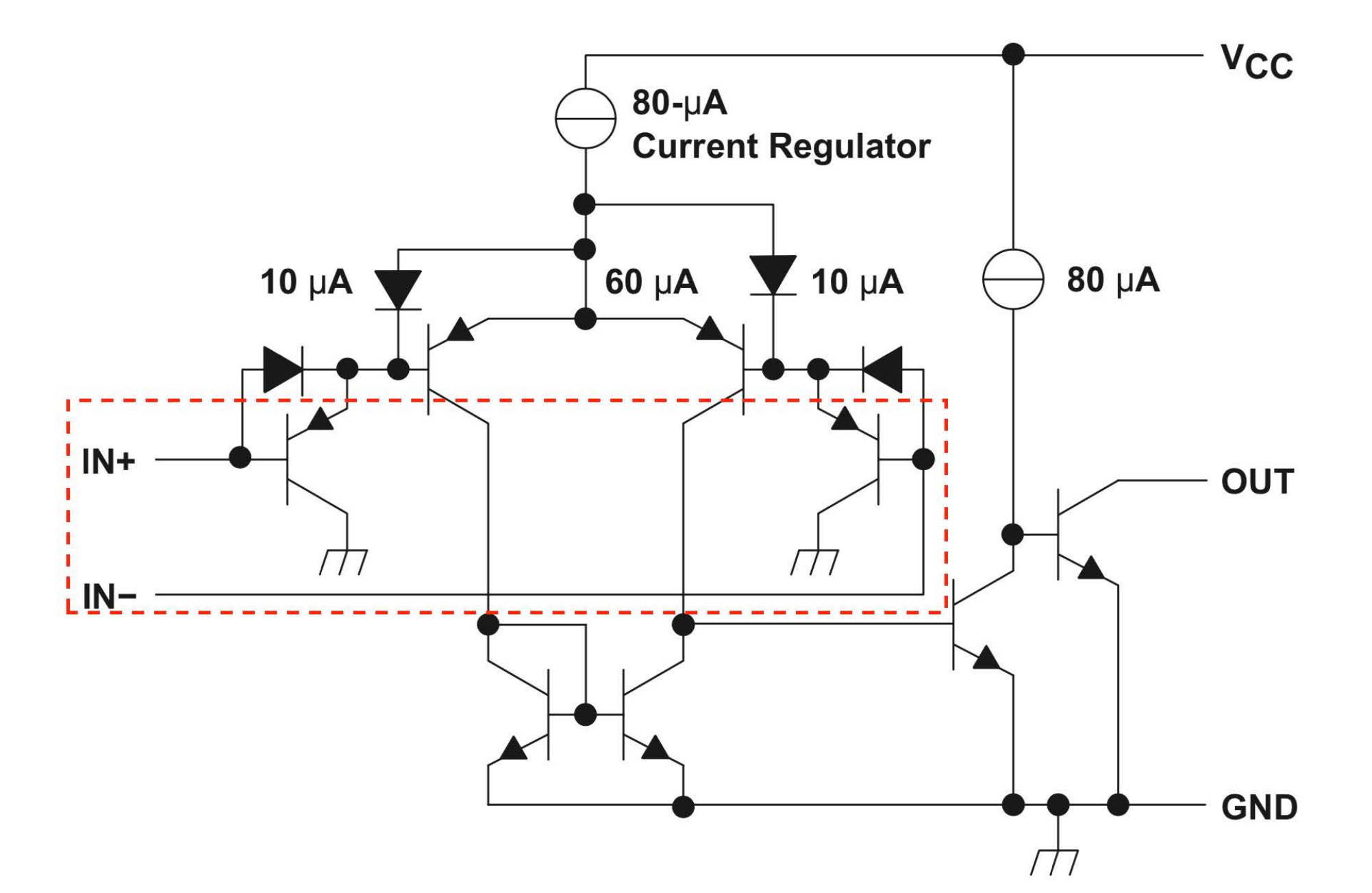} 
\caption{(color online)
Circuit schematic of quad comparators LM2901~\cite{LM2901_datasheet}.}
\label{fig:LM2901_Circuit}
\end{figure}

To investigate the variability of the irradiation-induced degradation in bipolar devices, one quad comparators, LM2901, and
two quad-operational amplifiers, MC4741 and LM324N, are used in our experiments.
These devices are chosen due to three reasons.
First, bipolar devices with simple input stages are chosen to avoid nonlinearity of TID response in devices with compensated input stages~\cite{barnaby1996analysis,dusseau2006analysis,Johnston2006_IEEETNS53-1779}.
The circuit schematic of LM2901~\cite{LM2901_datasheet} is shown in Fig.~\ref{fig:LM2901_Circuit}, from which a simple input stage can be seen.
Secondly, both comparators (LM2901) and amplifiers (MC4741 and LM324N) are chosen for generality.
Thirdly, both amplifiers with NPN-type (MC4741) and PNP-type (LM324N) input stages are chosen for generality.
For clearness, the characteristics of these devices and the related test conditions are summarized in Tables~\ref{tab:device} and II, respectively.

\begin{table}[!h]
\caption{Characteristics of the devices.}
\label{tab:device}
\centering
\begin{tabular}{ccccccccc}
\hline
Type      & Manufacture              & Function                & Input stage & Package     
\\
\hline
LM2901  & TI   &  Quad comparator & PNP           & PDIP (14)    
\\
MC4741  & Motorola                  &  Quad op amp        & NPN          & Plastic DIP 
\\
LM324N  &  TI  &  Quad op amp        & PNP          & PDIP (14)   
\\
\hline
\end{tabular}
\end{table}

\begin{table}[!h]
\caption{Test conditions of the devices.}
\label{tab:condition}
\centering
\begin{tabular}{ccccccccc}
\hline
Type      
& Test size & Dose rate  & Temperature 
\\
\hline
LM2901  
& 75$\times$4           & 0.1 rad/s    &   $25^{\circ}$C  
\\
 & 25$\times$4           & 100 rad/s   &   $25^{\circ}$C  \\
MC4741  
 & 75$\times$4           & 0.1 rad/s    &   $25^{\circ}$C  
\\
& 25$\times$4           & 100 rad/s   &  $25^{\circ}$C  \\
LM324N  
& 6$\times$4             & 10 rad/s    &   $25^{\circ}$C  
\\
& 4$\times$4             & 10 rad/s    &  $80^{\circ}$C  \\
& 4$\times$4            & 10 rad/s    &  $120^{\circ}$C  \\
& 4$\times$4             & 10 rad/s    &  $175^{\circ}$C  \\
\hline
\end{tabular}
\end{table}

{The experiments are carried out according to the standard test procedure, MIL-STD-883-G, Test Method 1019.7.}
For each type of device, large sample sizes are used to obtain credible results.
For LM2901 and MC4741, 25$\times 4$ (75$\times 4$) devices were chosen randomly from the same manufacturer and a same lot and irradiated with high (low) dose rate.
For LM324N, 6$\times 4$ (4$\times 4$) devices were randomly chosen from the same manufacturer and a single lot and irradiated at room (higher) temperature.
Every single device was labelled so that its parameter changes can be tracked.

The radiation source is a Cobalt-60 gamma ray source 
in Xinjiang Technical Institute of Physics \& Chemistry, Chinese Academy of Sciences.
Dosimetry was performed using an ionizing radiation absorption dose measurement system containing
a PTW 0.6cc Farmer ionization chamber.
The expanded uncertainty of the measurement is 4.052\% and the confidence coefficient is 95\%.

In order to irradiate multiple devices at the same time while ensuring identical levels of dose rate on each device, all 75$\times 4$ or 25$\times 4$ (6$\times 4$ or 4$\times 4$) devices
were closely placed on an irradiation board so as to lie on the same isodose panels.
The irradiation board measures $\rm 30 cm \times30 cm$ and the total dose difference at any point on it is no more than 5\% for all the chosen radiation dose rates.
The chips are irradiated with all pins grounded.

To consider the possible influence of irradiation dose, dose rate, and temperature {on the variability of the TID effects},
the devices LM2901 and MC4741 are irradiated at high dose rate (100 rad/s(Si)) and low dose rate (0.1 rad/s(Si)) field at room temperature,
while the devices LM324N are irradiated at room temperature and high temperature ($T=80$, 120, and 175 $^{\circ}$C) at 10 rad/s(Si) dose rate, respectively; see Tab. II.
{The elevated temperature is provided by an ESPEC heating system with accuracy of $\pm2^\circ$C.}
For LM2901 and MC4741, parameters are measured when the total dose is accumulated to 10 krad (Si), 30 krad (Si), 50 krad (Si), and 100 krad (Si).
For LM324N, the total dose is accumulated to 5 krad, 10 krad, 30 krad, 50 krad, and 100 krad.

Electrical parameters of the devices were measured at room temperature for pre- and post-irradiation. 
For the quad comparator LM2901N, we tracked the following parameters:
input offset voltage (Vio), IBC (Ib,+/-),
input offset current (Iio), open loop gain in dB (Avd), and output voltage (Vo).
For the quad operational amplifiers MC4741 and LM324N, we measured the following parameters:
Vio, Ib (+/-), Iio, Avd, and common-mode rejection ratio (Kcmr).
{Remote testing} were performed with an integrated parameter analyzer for amplifiers and comparator, SIMI3193.
The analyzer provides a test box and a test adapter to minimize the impact of leakage on the measurements.
The analyzer has a current resolution down to 0.1nA and an uncertainty less than 5\%.
The resolution is much smaller than the smallest pre-irradiation IBC ($\sim 10$nA).
Each test was finished within half an hour after irradiation to guarantee that the annealing process does not have a great impact on the result.

\section{Results and Discussion}
\label{sec:results}

Our results show that, for all three types of device,
the IBC is the most sensitive parameter.
So, we focus on the variability of this parameter here.
The pre- and post-irradiation IBC
are denoted by $I_B^0$ and
$I_{B}$, respectively. $\Delta I_{B} = I_{B} - I_B^0$ is the $\gamma$-ray induced ionization damage.

\begin{figure}[!t]
\centering
\includegraphics[width=0.7\linewidth]{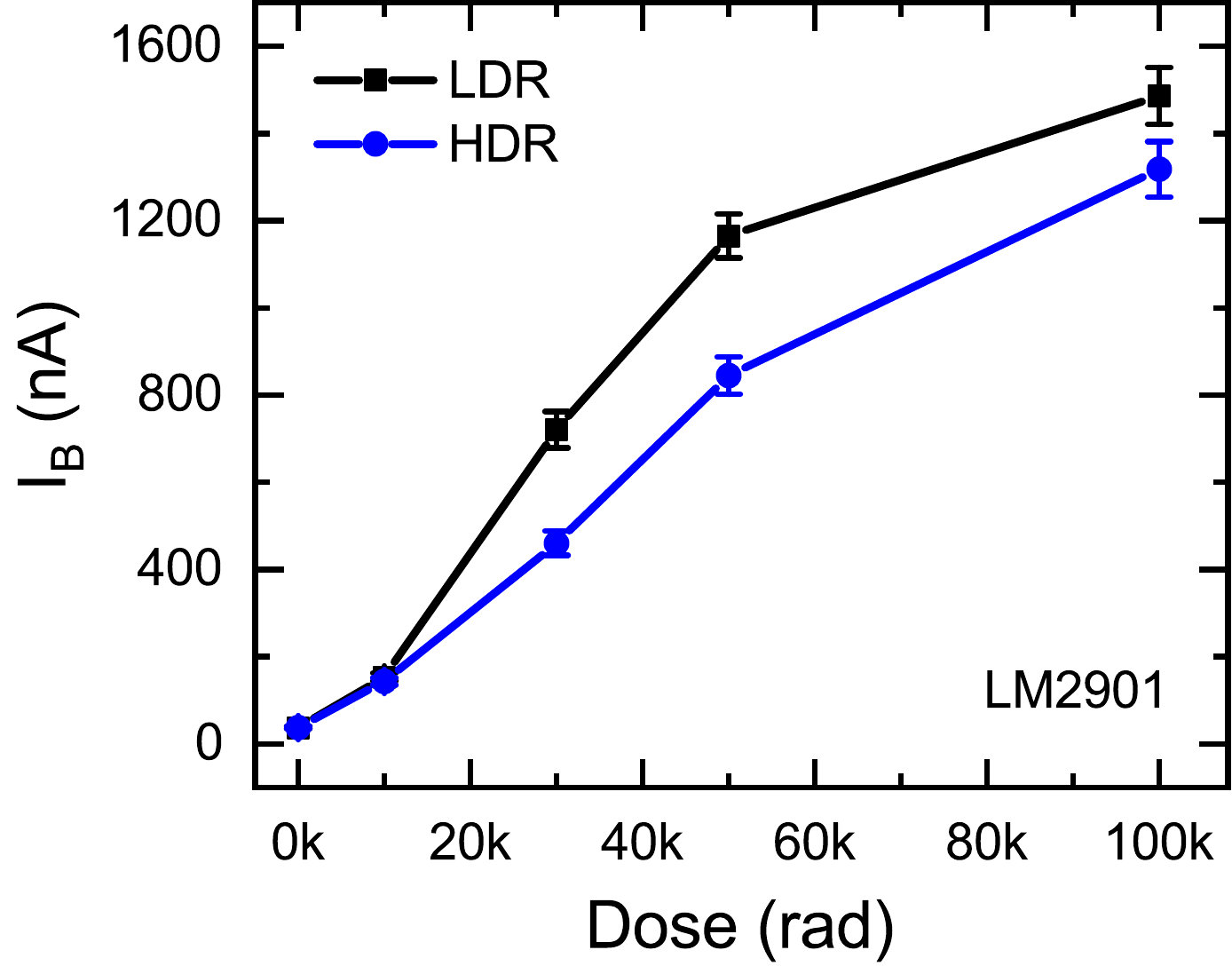} 
\caption{(color online)
{The average value and standard derivation of IBCs of comparator LM2901 as functions of the total dose.
Black (blue) for $75\times 4$ ($25\times 4$) comparators irradiated at low (high) dose rate. }
}
\label{fig:LM2901_Dose_IB}
\end{figure}

\subsection{Linear {$I_B^0$ dependence} in comparator LM2901}

The changes in the IBC of comparator LM2901 as function of the total dose
are shown in Fig.~\ref{fig:LM2901_Dose_IB}.
We can see clearly that 
{the mean value of} $I_{B}$ increases monotonically as the total dose increases.
{For each tested dose}, the observed IBC at low dose rate are always larger than that
of the high dose rate, indicating the presence of an enhanced low-dose-rate sensitivity (ELDRS) effect~\cite{Enlow1991_IEEETNS38-1342} for LM2901.
We also observe that as the total dose increases,
the current-dose curves of these samples do not crossover; instead, the differences of $I_{B}$ between samples  increase with increasing dose,
which leads to a \emph{spread} behavior of the data;
{see the increasing standard deviations (error bars) in Fig. 2.}
{This behavior was also observed in the input offset voltage of operational amplifiers LM324 and LM358~\cite{bakerenkov2019experimental}.}

\begin{figure}[!t]
\centering
\includegraphics[width=\linewidth]{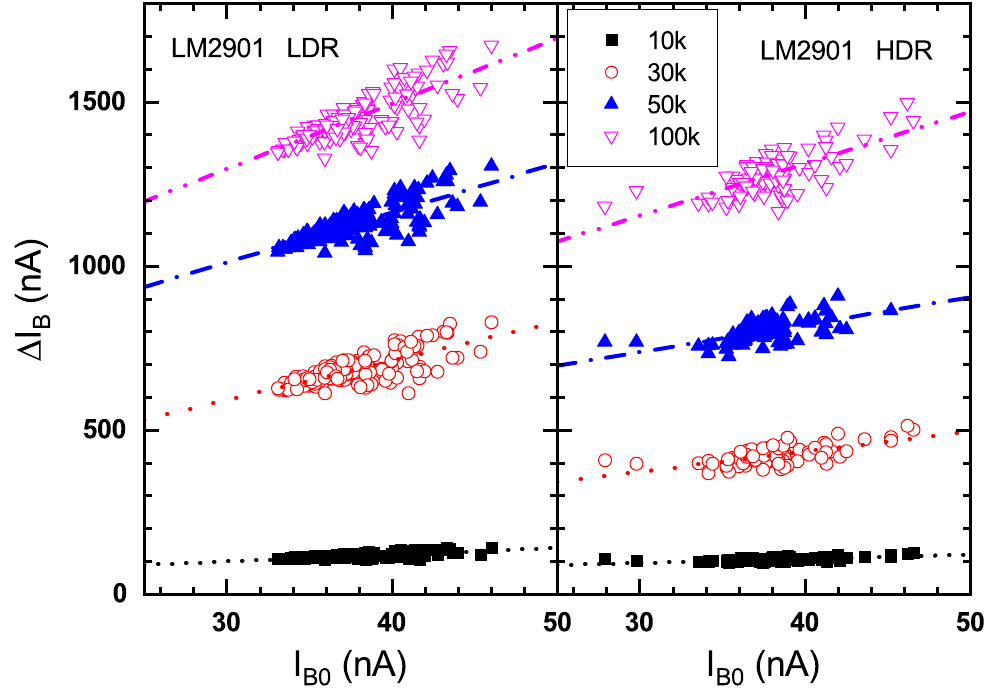}
\caption{(color online)
IBC's degradations versus the pre-irradiation values for comparator LM2901
irradiated at low (left panel) and high (right panel) dose rate.
The short-dashed, dotted, dash-dotted, and dash-dot-dotted lines represent the linear fitting
of the data at different total doses as indicated in the figure.
}
\label{fig:LM2901_DIB}
\end{figure}

\begin{figure}[!t]
\centering
\includegraphics[width=0.9\linewidth]{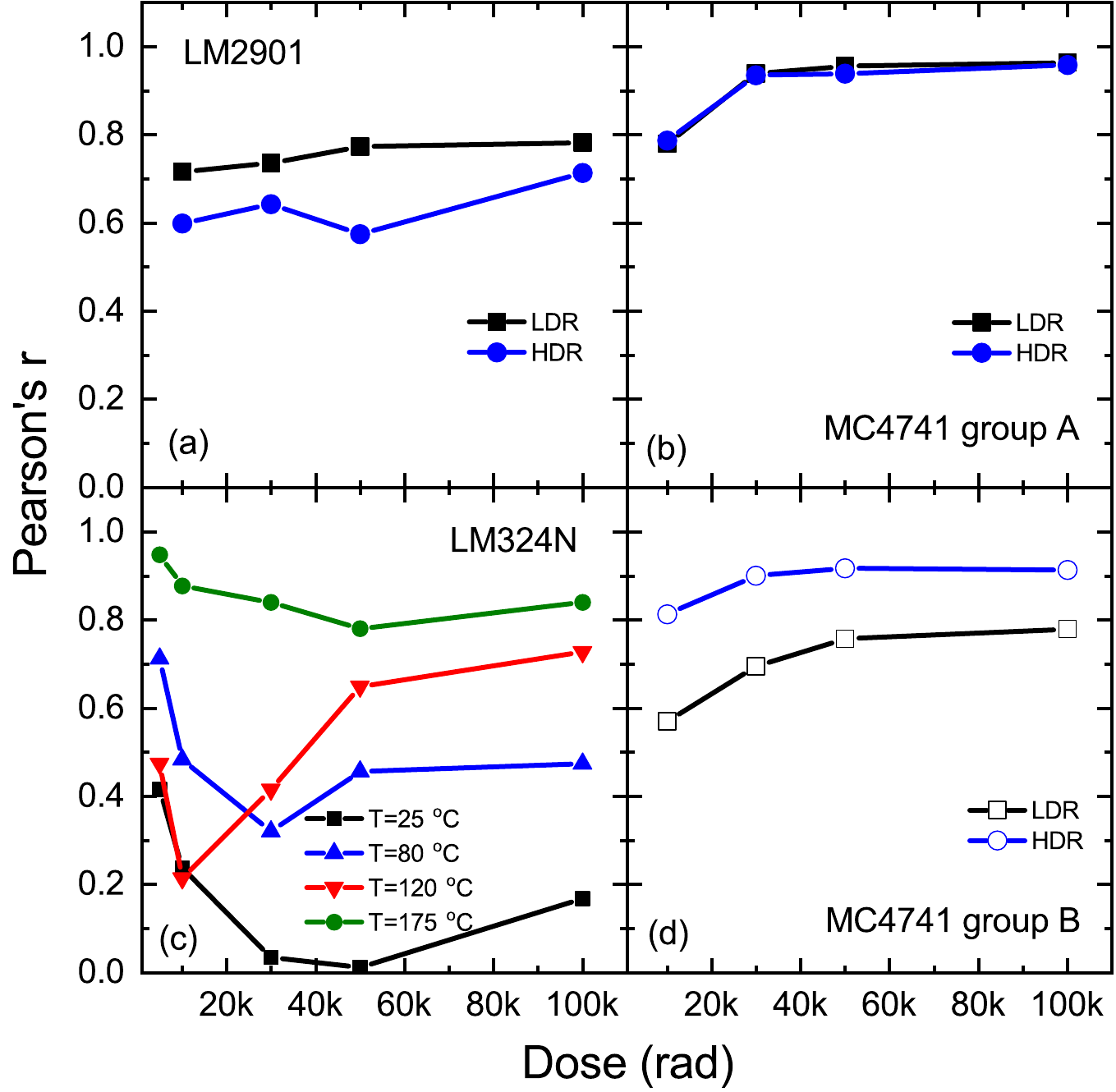} 
\caption{(color online)
Pearson's correlation coefficients {for the linear dependence} as a function of total dose for LM2901 (a),
group A of MC4741 (b), group B of MC4741 (d), and LM324N (c).
}
\label{fig:pearsonR}
\end{figure}

\begin{figure}[!t]
\centering
\includegraphics[width=0.97\linewidth]{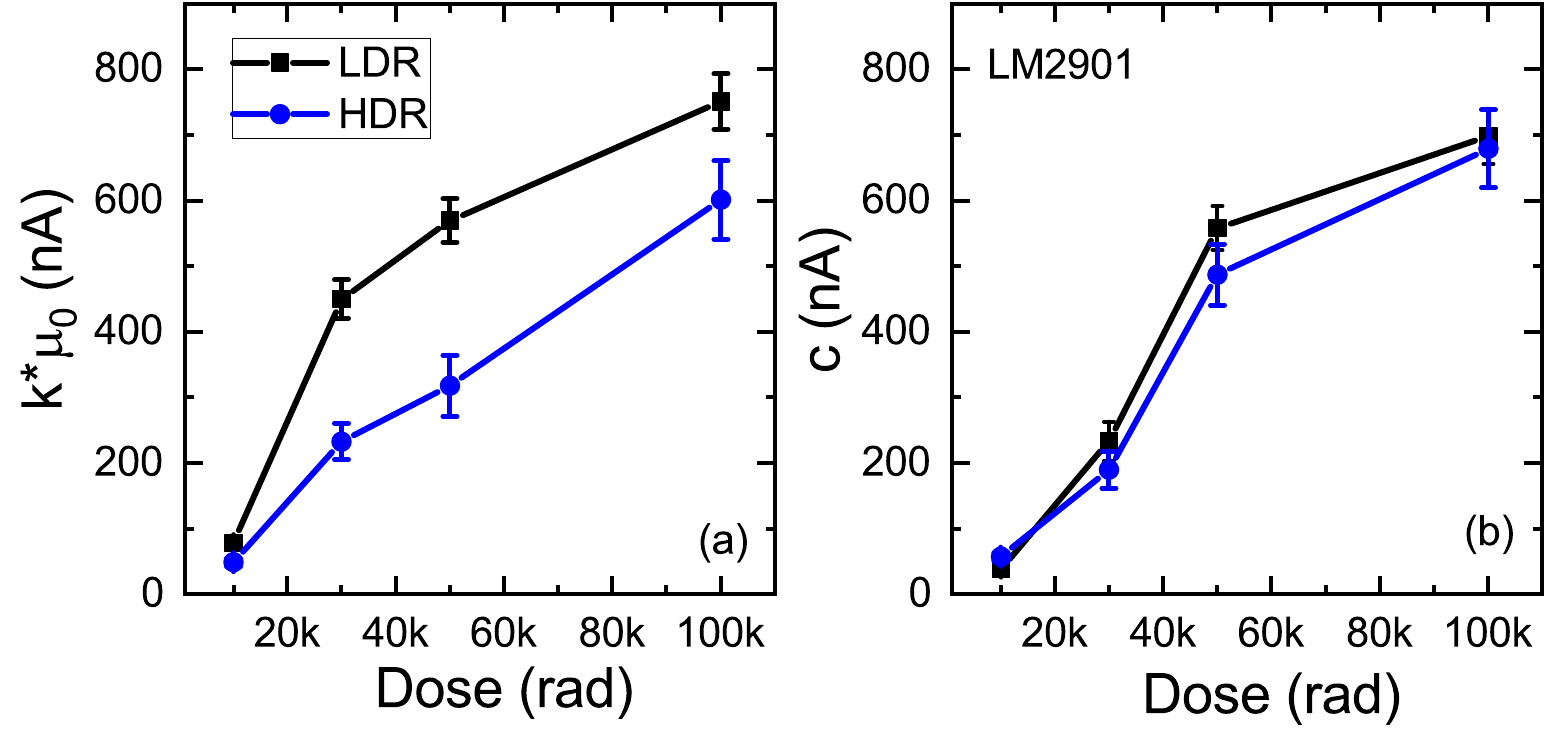} 
\caption{(color online) 
The linearly-dependent component $k \times \mu_{0}$ (a) and the independent component $c$ (b)
as functions of the total dose for comparator LM2901.
Black (blue) for low (high) dose rate.
}
\label{fig:LM2901_slope}
\end{figure}

\begin{figure}[!b]
\centering
\includegraphics[width=\linewidth]{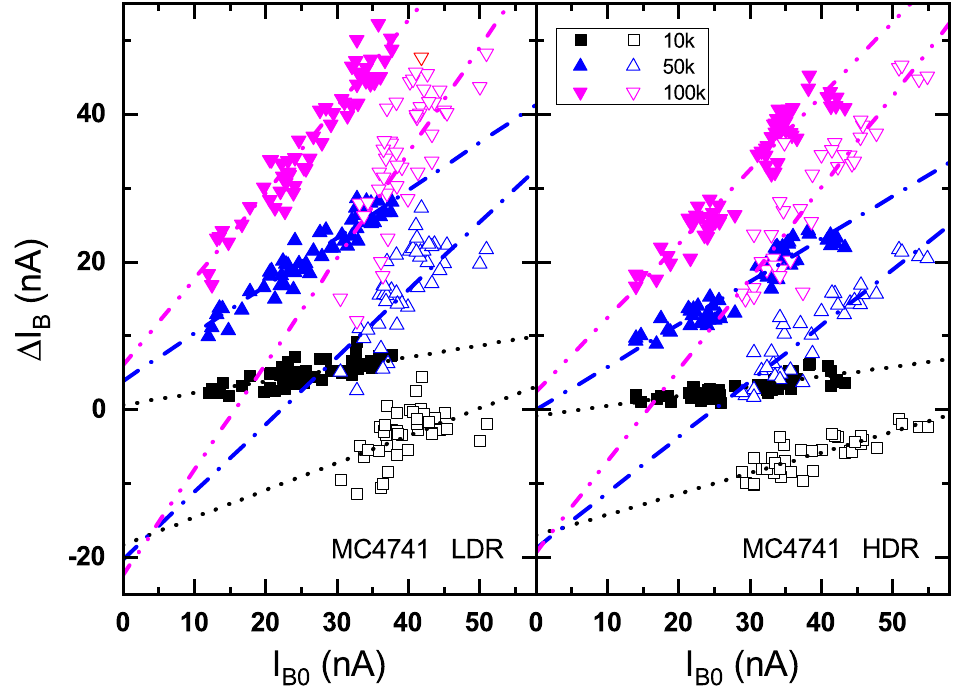} 
\caption{(color online)
Degradations of IBC versus the pre-irradiation IBC for amplifier MC4741
irradiated at low (left panel) and high (right panel) dose rate.
The data are divided into two groups due to their different responses.
The lines represent the linear fitting of the data at different total dose as indicated in the figure.
}
\label{MC4741_DIB}
\end{figure}

Does this interesting behavior imply a deep relation between the post-irradiation IBC and the pre-irradiation ones?
To testify this speculation, in Fig.~\ref{fig:LM2901_DIB}
we make a post-irradiation vs. pre-irradiation (P-P) plot
to describe the damage of all tested samples.
Remarkably, it is seen that,
the whole trend of the damage follows a simple linear dependence on the pre-irradiation values.
In other words, the samples with larger pre-irradiation IBCs will degrade more.
Moreover, from the figure, the linear dependence maintains for any dose and for both the low and high dose rates.
To further confirm the dependence, the Pearson's correlation coefficient is calculated and plotted in Fig.~\ref{fig:pearsonR} (a).
From the results, we can confirm {a linear dependence of the irradiation response on the initial value of IBC;
moreover, it is seen that the higher the irradiation dose, the more significant the linear dependence.}

The $\Delta I_{B}$-$I_B^0$ data can be fitted using a {linear regression model}:
\begin{equation}
\Delta I_B (I_B^0) = k \times I_B^0 + c,
\label{eq:linear}
\end{equation}
where $k$ is the slope of the linear dependence and $c$ is a constant component independent of the pre-irradiation IBC.
The linearly-dependent component of $\Delta I_B$ then can be calculated as $k\times\mu_{0}$, where $\mu_{0}$ is the mean value of $I_B^0$.
The results are plotted in Fig.~\ref{fig:LM2901_slope}, from which we obtain that both
the linear and constant component increase for increasing dose at a fixed dose rate
and both increase for decreasing dose rate at a fixed dose {that is larger than 10 krad (Si)}.

\subsection{Linear {$I_B^0$ dependence} and bimodal response in OA MC4741}

\begin{figure}[!t]
\centering
\includegraphics[width=\linewidth]{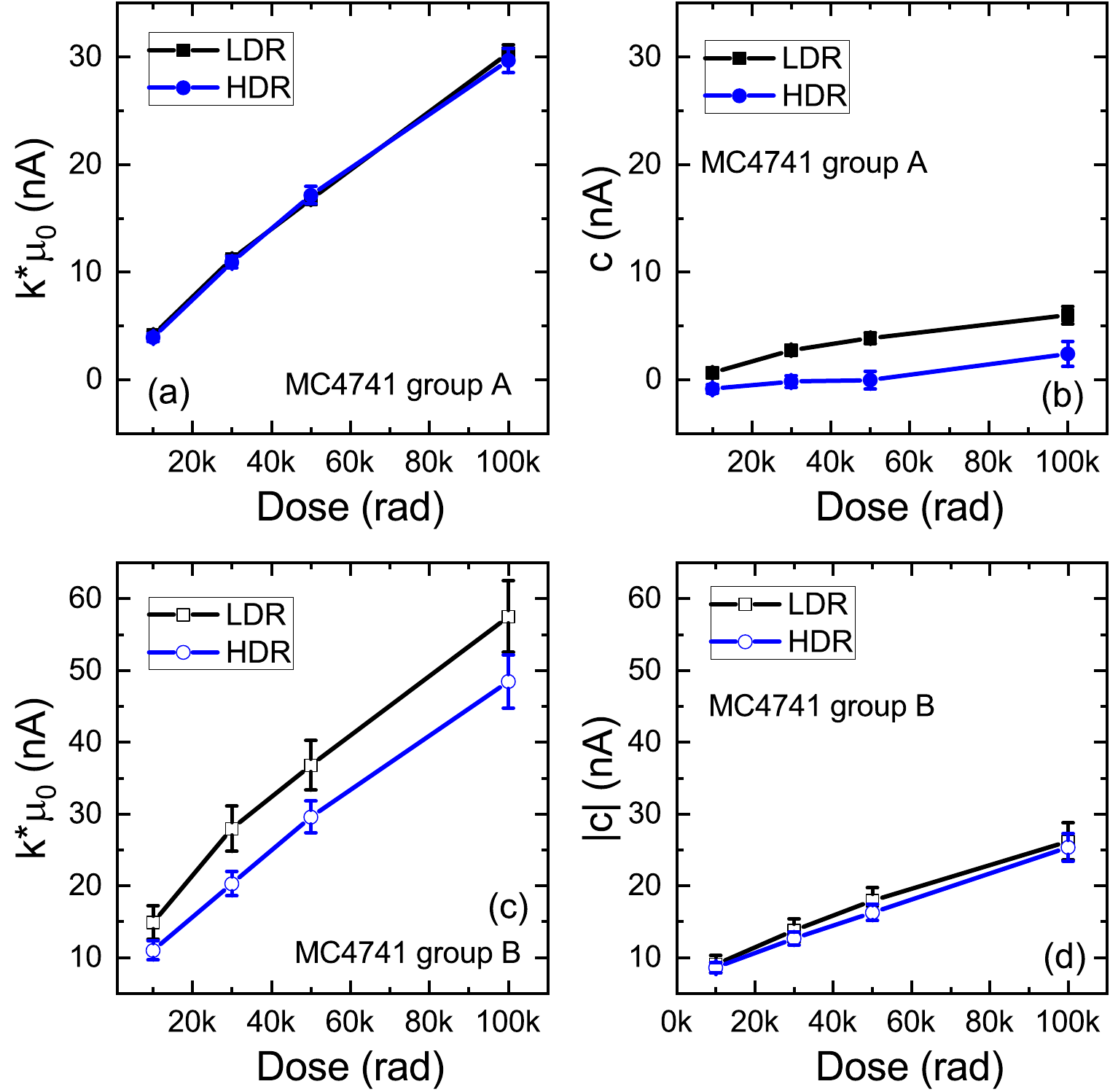} 
\caption{(color online)
The linearly-dependent component $k \times \mu_{0}$ (a, c) and the independent component $c$ (b, d)
as functions of the total dose for amplifier MC4741.
The solid (open) symbols represent the data for group A (B) as explained in the caption of Fig.~\ref{MC4741_DIB}.
Note, in (d) the current is negative and plotted in its absolute value.
}
\label{MC4741_slope}
\end{figure}

One may wonder whether the observed {linear dependence} be specific for the investigated device.
To this end, we investigate another bipolar device, operational amplifier MC4741.
Its input stage is also simple but the input transistor is NPN type.
The {P-P} plot is present in Fig.~\ref{MC4741_DIB}.
{A linear $I_B^0$ dependence is also observed for $\Delta I_{B}$ of this device.}
Actually, the Pearson's correlation coefficients (see Fig.~\ref{fig:pearsonR} (b)) is even larger than those for LM2901.
Similar to LM2901, {the linear dependence becomes more significant} for increasing dose.
{The linearly-dependent and independent components are extracted using Eq. (1) and plotted in Fig.~\ref{MC4741_slope} (a) and (b).
It is seen that, while the two components are comparable for LM2901, here the dependent current is much larger than the independent current in MC4741.}

\begin{figure}[!t]
\centering
\includegraphics[width=0.75\linewidth]{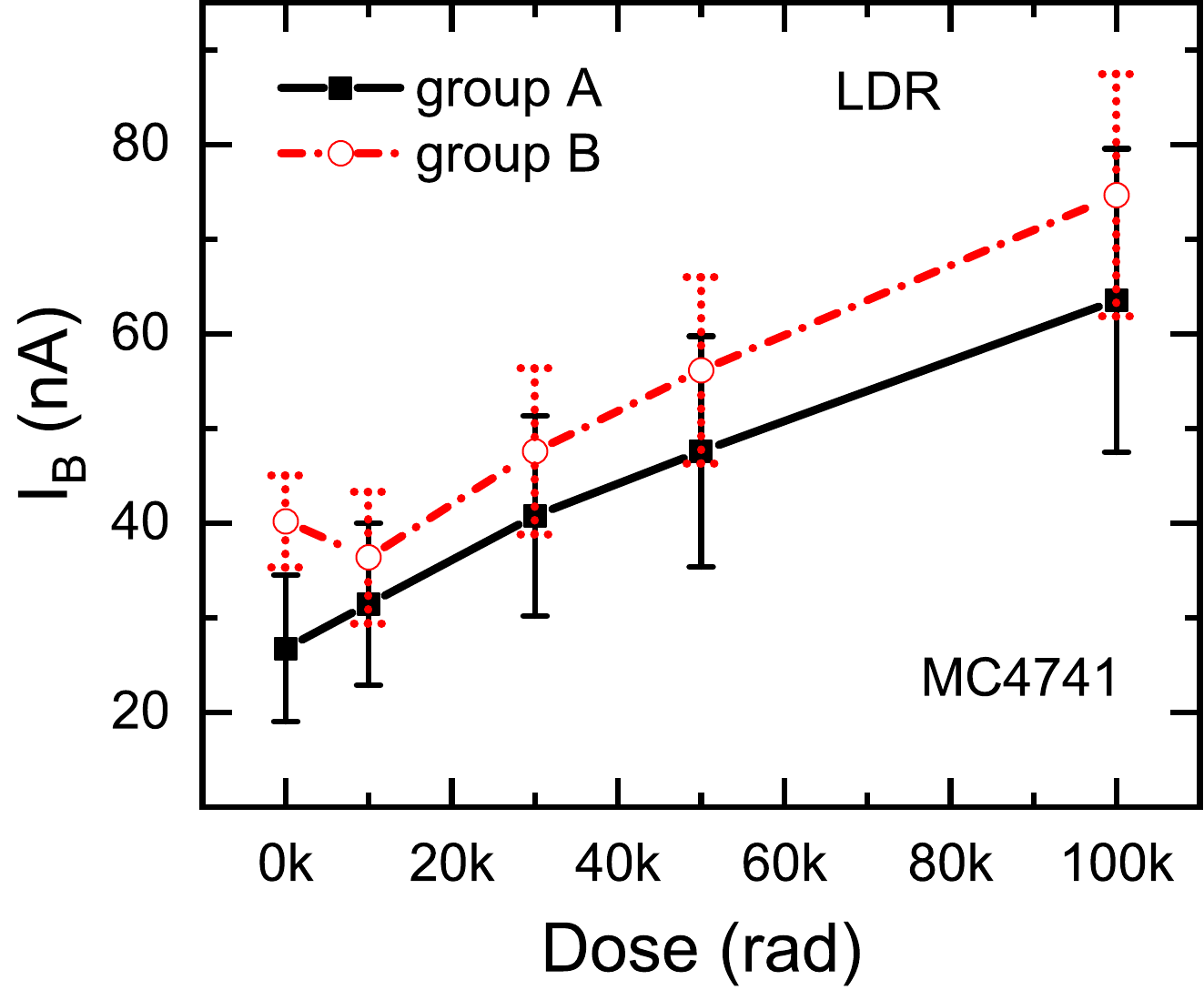} 
\caption{(color online)
{The average value and standard derivation of IBCs as functions of the total dose for amplifier MC4741, irradiated with low dose rate. Black (red) for amplifiers from the group A (B). }
}
\label{MC4741_LDR}
\end{figure}

From Fig.~\ref{MC4741_DIB}, we have another important observation:
the irradiation response of MC4741 is bimodal. 
The bimodal response is that, for any doses, there is a group (group B) which has
the similar slopes as the normal group (group A) but negative intercepts,
see Fig.~\ref{MC4741_slope} (c) and (d), respectively. 
For the Pearson's correlation coefficients, see Fig.~\ref{fig:pearsonR} (d).
The bimodal response was also observed in the IBC of LM111~\cite{Krieg1999_IEEETNS46-1627} {and the input offset voltage of LM358~\cite{bakerenkov2019experimental}.}
To investigate the possible reason, {we plot the current-dose profile for each 
group in Fig.~\ref{MC4741_LDR}.
{For group A, we see clearly a spread behavior of the damage similar to Fig. 2:
the larger the dose the bigger the standard deviations.}
For group B, besides the spread behavior, the average value of IBCs 
first decreases and then increases with the increasing gamma ray dose.}
{As the input offset voltage in LM358 is a complex function of the total dose for each type of the input stage, the authors of Ref.~\cite{bakerenkov2019experimental} connected the bimodal behavior of it to the "circuit" factor, i.e., the asymmetry of a current mirror of the differential stage as well as the difference of I-V characteristics of differential pair transistors.
Here, the IBC of the devices are directly
related to the base current of the input-stage transistors~\cite{Pease1996_IEEEREDW-28,Barnaby1999_IEEETNS46-1666,
dusseau2006analysis,Krieg1999_IEEETNS46-1627}.
On the other hand, from Fig. 8 we notice that, the devices of group B have a larger initial IBCs than those of devices of group A.
So, there should be pre-existing defects in silicon of group B devices due to the process of ion implantation, and the abnormal decrease behavior of IBCs of group B devices is an annealing of these defects under ionization irradiation.
Such an annealing is the so-called injection annealing~\cite{gregory1965injection}, in which the reordering of pre-existing defects is enhanced by the presence of free charge carriers induced by ionization irradiation, due to the change in defect charge states and enhancement of defect mobility~\cite{kimerling1978recombination,song2020defect,song2020origin}.}
{Nevertheless, although the MC4741 device response in a bimodal way,
the post-irradiation IBCs still {linearly depend on} the pre-irradiation ones.}

\begin{figure}[!t]
\centering
\includegraphics[width=\linewidth]{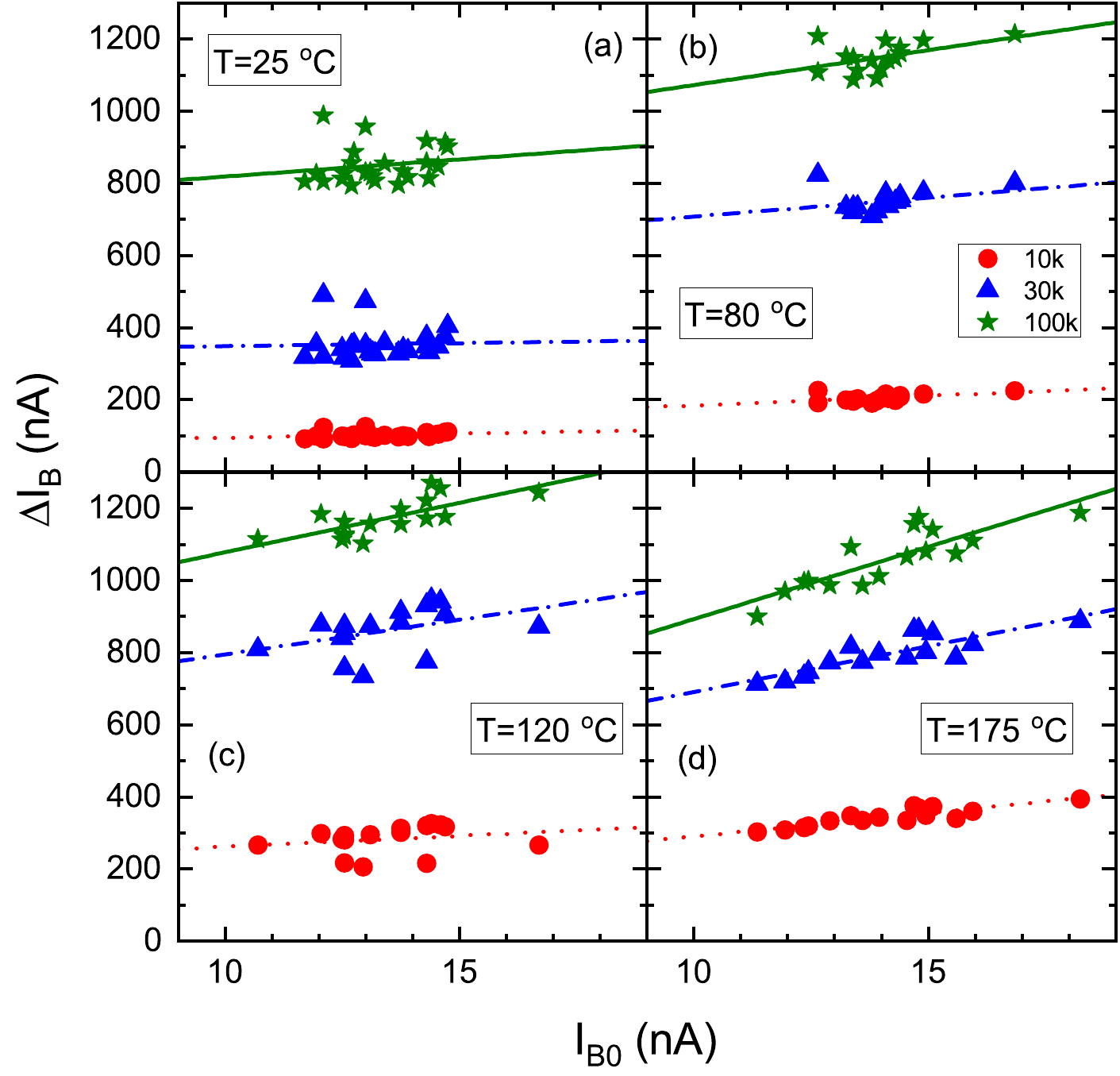} 
\caption{(color online)
Degradations of the IBC versus the pre-irradiation IBC for amplifiers LM324N.
The lines represent the linear fitting of the data at different total dose as indicated in the figure.
The experiments are performed at room temperature $T=25$ $^{\circ}$C (a) and at high temperatures
with $T=80^{\circ}$C (b), 120$^{\circ}$C (c), and 175 $^{\circ}$C (d), respectively.
}
\label{LM324N_DIB}
\end{figure}

\begin{figure}[!b]
\centering
\includegraphics[width=0.97\linewidth]{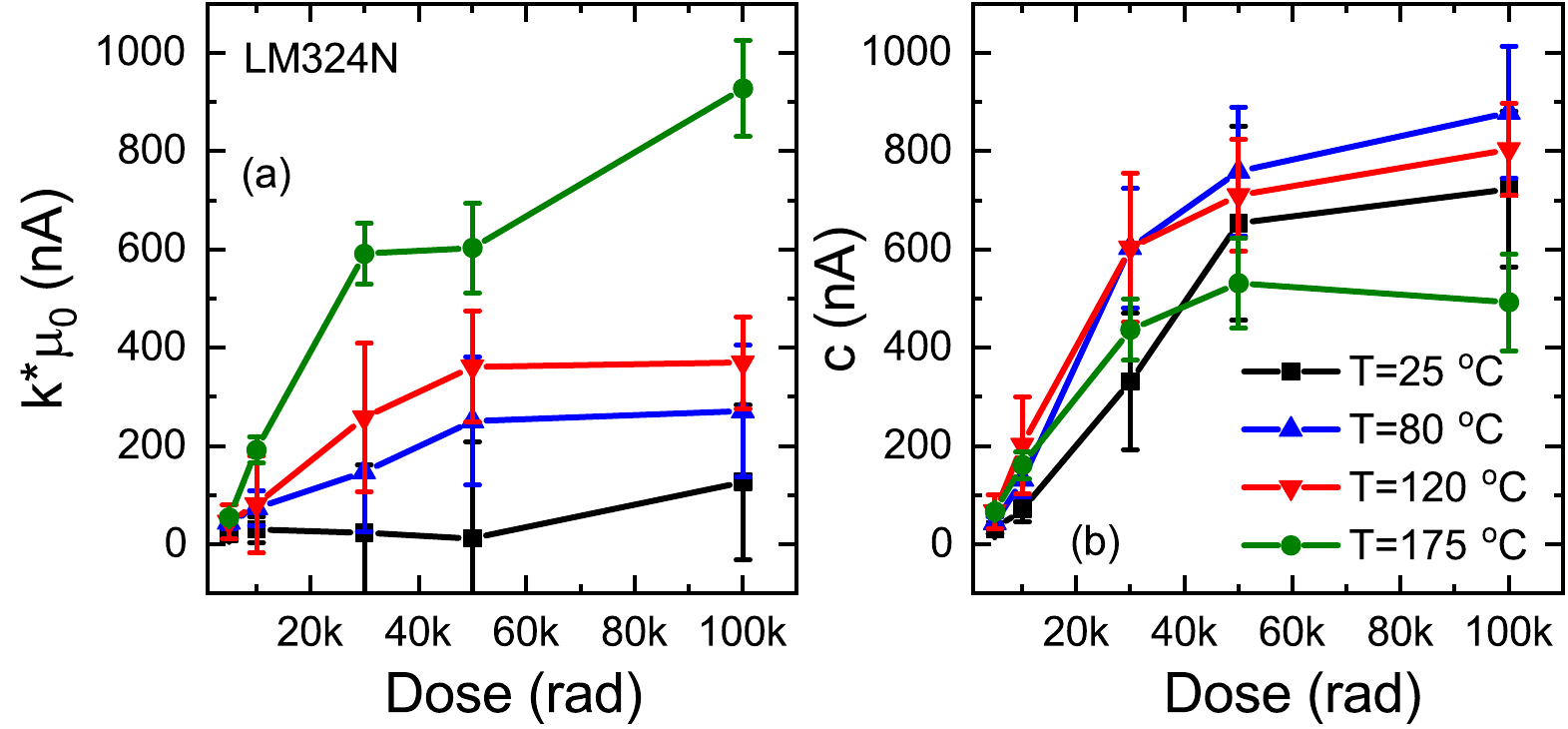} 
\caption{(color online)
The linearly-dependent component $k \times \mu_{0}$ (a) and the independent component $c$ (b)
as functions of the total dose for amplifier LM324N.
}
\label{LM324N_slope}
\end{figure}

\subsection{{Linear $I_B^0$ dependence} in LM324N at high temperature}

The experiments for comparator LM2901 and amplifier MC4741 are performed at room temperature.
Would the {linear dependence} hold for high-temperature irradiation?
To check this possibility,
we performed the gamma ray irradiation experiments for amplifier LM324N (with PNP-type simple input stage)
at both room temperature ($T=25$ $^{\circ}$C)
and high temperatures of $T=80$, 120, and 175 $^{\circ}$C.
The dose rate is set to 10 rad(Si)/s.
The P-P plot of obtained results are shown in Fig.~\ref{LM324N_DIB}.
It is seen that the {linear $I_B^0$ dependence of $\Delta I_B$} also holds for these higher temperatures.
This is clearly reflected by the Pearson's correlation coefficient, which is plotted
in Fig.~\ref{fig:pearsonR} (c) as a function of the total dose for different temperatures.
Different from the other two devices, the coefficient first decreases and then increases with increasing dose.
On the other hand, the higher the temperature, {the more significant the linear dependence}.

The two derived current components using Eq. (1) are displayed in Fig.~\ref{LM324N_slope}.
It is seen that the {linearly-dependent} current component is much sensitive to the elevated temperature than the independent current component.
At room temperature, the independent current component dominates;
at higher temperature, the {dependent} current component becomes comparable or even larger.

\begin{figure}[!t]
\centering
\includegraphics[width=0.87\linewidth]{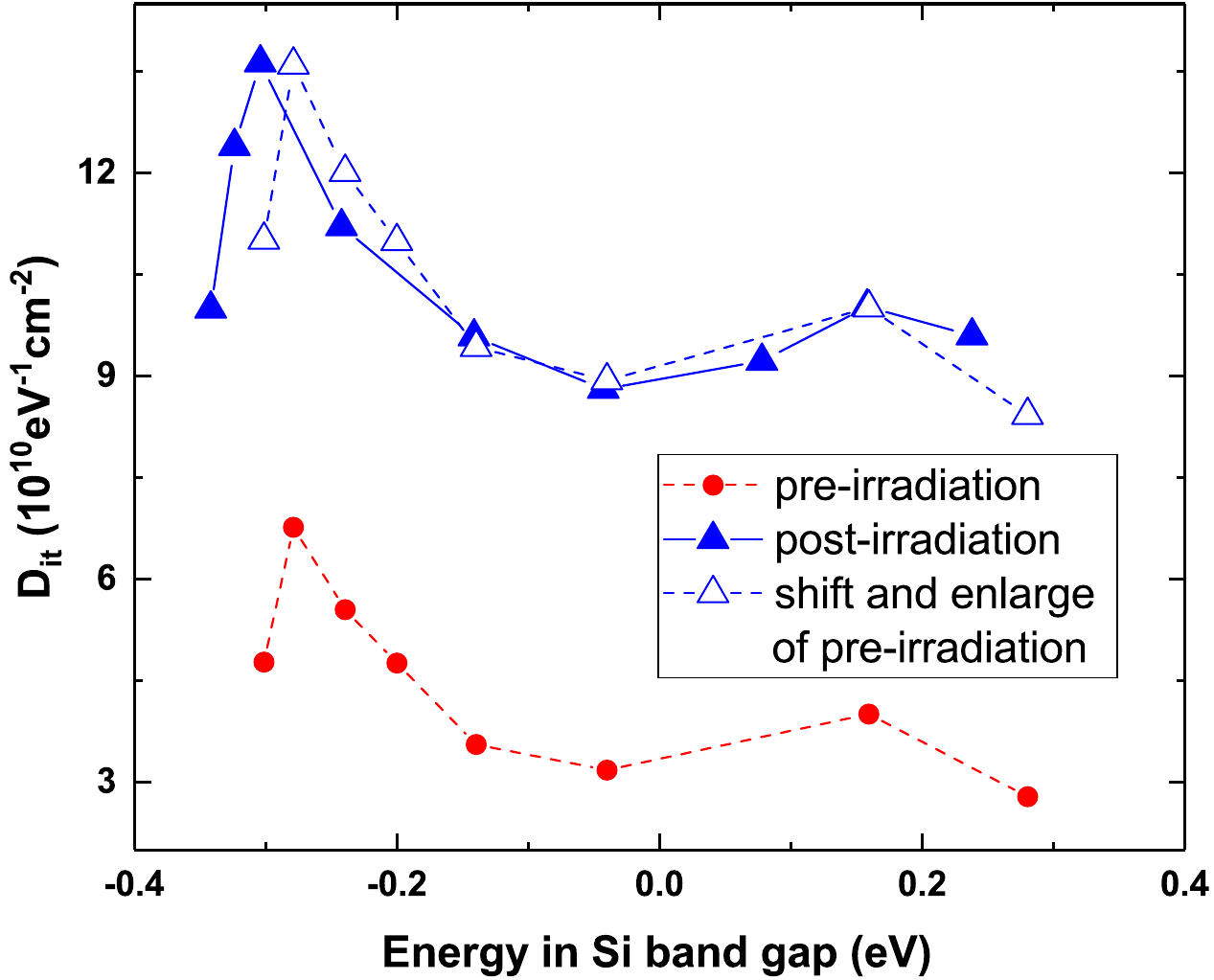} 
\caption{{(color online)
The interface state distribution in the energy gap of Si before (red dashed) and after (blue dashed) ionization irradiation. As shown by the blue solid curve, the latter is found to be an overall shift and enlarge of the former,  $D_{it}(E)=1.3\times D_{it}^0(E) + 4.8$. The data are adopted from Ref.~\cite{ma1975comparison}.
}
}
\label{fig:BJT}
\end{figure}

\subsection{Possible Mechanism of the observed {linear dependence}}

{The linear dependence of $\Delta I_B$  on $I_{B0}$ is observed in devices of different manufacturers, different structures, and different irradiation conditions (see Tabs. I and II). These facts 
suggest that the {linear dependence} is a rather general phenomenon
that is independent to the specific process and structure of the device.}
In following, we will suggest that such a general phenomenon may stem from 
{a unique behavior of the energy distribution of the interface states ($D_{it}$) under ionization irradiation.
{There are three reasons.}
Firstly, it is noticed that the {linear dependence} are observed even at very small total dose.
As indicated in  Refs.~\cite{Pease1996_IEEEREDW-28,Barnaby1999_IEEETNS46-1666,dusseau2006analysis,Krieg1999_IEEETNS46-1627},
in this situation the IBC of the 
devices are directly related to the base current of the input-stage transistors.
Secondly, the base current is determined by the surface recombination velocity, which is a product of the caption cross-section and the density of interface states in the oxide above the base region~\cite{pierret1987advanced}.
{The caption cross-section is almost unchanged under irradiation, hence, the irradiation-induced change in $D_{it}$  can dominate the origin of the linear dependence between $\Delta I_B$  and $I_{B0}$. 
Thirdly and lastly, we do find a linear dependence of the post-irradiation interface state distribution on the pre-irradiation one, see below for details. }

As observed in experiments~\cite{sivo1972investigation,ma1975comparison}, 
the pre-irradiation interface state distributes nonuniformly in the energy gap of silicon, see the red dots in Fig. 11.
Due to the variability of the structure parameters of the base region, the input-stage transistors of the same manufacturer and a single lot can have different effective Fermi levels in the base, 
at which the densities of interface state are different.
This leads to different interface recombination velocities, hence different base currents of the transistors and different IBCs of the devices.}

{
Under ionization irradiation, new interface traps will be generated due to the depassivation of Si-H bonds by protons at the interface~\cite{saks1989interface,rashkeev2001defect}, see the blue dots in Fig. 11.
The protons are released in the bulk silica through the interaction of irradiation-induced oxide trapped charges and hydrogen~\cite{stahlbush1990post}. 
Interestingly, the generated interface state displays a similar distribution as the pre-irradiation $D_{it}$~\cite{sivo1972investigation,ma1975comparison}, see Fig. 11.
The reason is that, a part of the 
Si-H bonds is located near the initial interface traps and has a similar physical environment~\cite{ma1975comparison}. 
Hence, at any fixed energy, the larger the initial $D_{it}$, the larger the generated $D_{it}$ under certain irradiation dose. 
As a result, devices with larger IBCs, i.e., with larger interface states near their Fermi energies, will have larger increase of interface states after certain irradiation, hence have larger increase of IBCs. 
This is the observed {linear $I_B^0$ dependence of the $\Delta I_B$}.

The irradiation-induced $D_{it}$ was believed to be proportional to the pre-irradiation $D_{it}$~\cite{sivo1972investigation,ma1975comparison}.
However, our calculation shows that the change is actually an overall shift plus an enlarge of the initial distribution, see the dashed curve in Fig. 11, which shows a higher position and an enhanced peak-to-valley ratio relative to the initial distribution profile.
The overall shift happens because the other part of Si-H bonds is far away from the initial interface traps, and the generated interface states should display an uniform energy distribution that has no {relation with} the initial $D_{it}$.
The total change of $D_{it}$ induced by irradiation can be expressed as 
\begin{equation}
\Delta D_{it}(E)=k'\times D_{it}^0(E) + c'.
\end{equation}
As the initial and generated $D_{it}$ near the Fermi energy corresponds to the  pre-irradiation and post-irradiation IBC, respectively,
{a linear dependence} of the same form should be observed for the IBC of the devices.
This is just Eq. (1) and the results as obtained in Figs. 3, 6, and 9.
It is noticed that, the irradiation-induced enlarge of the $D_{it}$ is indispensable for the linearly-dependent component in Eq. (1), while the irradiation-induced overall shift of the $D_{it}$ is response for the  independent component in Eq. (1).
As the dose increases, both the enlarge and shift of the $D_{it}$  increases.
This corresponds to the dose dependence of the linearly-dependent and independent components observed in Figs. 5, 7, and 10.}

As mentioned above, the linear component is dominant in
MC4741, while the constant component becomes dominant in LM324N.
{These results can be explained self-consistantly based on the above picture.
In MC4741, it is seen that the irradiation-induced increase of IBC is comparable with the initial value of IBC.
This fact implies that most of the interface traps are genearted through the Si-H bonds around the inital interface traps.
As a result, the enlarge of $D_{it}$ dominates the irradiation response and one observes a dominating linear component in the IBC of this device.
In LM324N, the irradiation-induced increase of IBC is much larger than the initial value of it.
This fact reflects that plenty interface traps are generated away from the initial ones. So, the shift of the $D_{it}$ dominates the response process and a larger independent component is observed in the IBC of this device.}

\subsection{Damage Prediction Using the {Dependence Phenomenon}}

In practice, the irradiation degradations of the IBC of to-be-use devices are usually estimated by
the mean damage of randomly selected samples from the same lot.
Here, we indicated that, 
this method can be inapplicable when
{the linearly-dependent component} is comparable with or even larger than {the independent component} (e.g., the LM2901 and MC4741 cases).
This is because, in these cases samples with small difference in $I_B^0$ (several nA)
can response very differently to the same irradiation (hundreds of nA), see Figs. 3 and 6.
As a result, the damages can be very different for each sample to be used and each sample has been tested.
For example, in the case of MC4741, the maximal damage can be 2.5 times of the minimal damage.
Here, based on the observed {linear dependence}, we propose that the damages of to-be-used samples can be more precisely predicted
by using their pre-irradiation IBCs as an index.
Specifically, using Eq.~(\ref{eq:linear}), the damages can be predicted by the pre-irradiation currents times the slope and plus the intercept;
the slope and the intercept can be obtained by carrying out large-size sample irradiation experiment
and extracting parameters from the P-P plot of the data.
It should be indicated that the proposed method also become inapplicable
for devices displaying bimodal response.
Of course, when {the independent} component dominates (the LM324N case), the traditional prediction method can be used to estimate the damages.

\section{Conclusion}

In summary, we have found a general experimental phenomenon in the irradiation response of COTS bipolar devices with simple input stages.
It is demonstrated that for devices from the same manufacturer and the same lot, samples with larger pre-irradiation IBCs will statistically degrade more under the same irradiation.
This {linear dependence} widely exists in all investigated cases, despite of the huge discrepancy in the device types, bimodal response, and irradiation conditions (dose, dose rate, and temperature).
We indicate that such a general phenomenon may stem from the unique behavior of the energy distribution of the interface states under irradiation.
The overall shift of the pre-irradiation $D_{it}$, which is due to the generation of interface traps away from the initial ones, is indispensable for the independent component of the post-irradiation IBC.
The enlarge of the pre-irradiation $D_{it}$, which stems from the generation of interface traps located near the initial ones, is responsible for the linearly-dependent component of the post-irradiation IBC.
Due to this mechanism, the {linear $I_B^0$-dependence} can be expected in other bipolar devices with simple input stages.
Our results also show that, 
the irradiation-induced linearly-dependent current can be comparable with the independent current in some bipolar devices.
In this case, the traditional average-value damage prediction method can be inapplicable.
Instead, we propose that the damage of to-be-use devices can be more precisely predicted by using the observed {dependence phenomenon} and the pre-irradiation currents.

\section*{Acknowledgments}
{We owe to an anonymous reviewer, without whose suggustions we will not connet the observed {linear dependence} with the energy distribution of the interface states.}
We also thank Dr. B.S. Tolleson of the Arizona State University and
Prof. D.M. Fleetwood of the Vanderbilt University
for insightful suggestions.
This work was supported by the Science Challenge Project (Grant No. 
TZ2016003-1) and National Natural Science Foundation of China (Grant Nos. 11804313 and 11404300).

\end{document}